\documentclass[showpacs,a4]{iopart}
\usepackage{psfig,epsfig,graphicx,float,pst-all,amssymb}
\renewcommand{\=}{~=~}
\newcommand{\be}{\begin{equation}}
\newcommand{\ee}{\end{equation}}
\newcommand{\bea}{\begin{eqnarray}}
\newcommand{\eea}{\end{eqnarray}}

\newcommand{\bc}{\begin{center}}
\newcommand{\ec}{\end{center}}
\begin{document}

\title{Analytically solvable potentials for $\gamma$-unstable nuclei.}
\author{Lorenzo Fortunato\footnote{email: fortunat@pd.infn.it} 
and Andrea Vitturi}
\address{Dipartimento di Fisica ``G.Galilei'' and INFN, via Marzolo 8, 
I-35131 Padova, Italy.}

\begin{abstract}
An analytical solution of the collective Bohr equation with a Coulomb-like
and a Kratzer-like $\gamma-$unstable potential in quadrupole deformation 
space is presented. Eigenvalues and eigenfunctions are given in closed form 
and transition rates are calculated for the two cases.
The corresponding SO(2,1)$\times$SO(5) algebraic structure is discussed.  
\end{abstract}
\pacs{21.60.Fw, 21.10.Re}
\maketitle

\section{Introduction}
Large interest has been recently raised by the analytic solution of the
$\gamma-$unstable collective Bohr hamiltonian in the case of a square well
potential in the $\beta$ variable \cite{Iac1}. This situation describes the 
shape phase transition between spherical and $\gamma-$unstable nuclei and it 
has been associated with the E(5) group. Similarly an approximate separation
of variables has been proposed \cite{Iac2} as a solution for 
the transition between spherical and axially deformed shapes which has been
referred to as X(5).
Experimental confirmation of the actual occurrence of such 
situations was found for instance in the 
examination of the level scheme of $~^{134}$Ba for E(5) and $~^{152}$Sm 
for X(5) \cite{Cas1,Bizz}.

The occurrence of dynamic symmetries is associated with systems where 
the hamiltonian is solely written in
terms of Casimir operators of a chain of subalgebras. The wider algebra of
this chain is the one that settles the nature of the problem and its general
form bears all the information on the system in a compact way.
The non-unique way in which the algebra is split in chains of subalgebras
exhibits some peculiar feature of the system leading to exact analytical
results, that may help in elucidating experimental findings and data trends.
Concerning ourselves to nuclei, 
an example of occurrence of dynamical symmetries is provided by the 
Interacting Boson Model. With particular choices of the general SU(6)
hamiltonian one obtains limiting situations that are analytically solvable 
(O(6),U(5) and SU(3) cases). Moving around in the nuclear chart or 
along a chain of isotopes, nuclei
can undergo phase transitions from one dynamical symmetry to another.
The interest of E(5) and X(5) lies in the fact that they are related to new 
analytically solvable situations corresponding precisely
to the critical phase transition points.

In the specific case of the transition 
from spherical to $\gamma-$unstable nuclei, the E(5) description 
assumes, for the transition
potential in the $\beta$ variable, an infinite square-well potential 
\cite{Iac1}, a case 
that has been generalized by Caprio with the introduction of a finite square
well \cite{Cap}.
Other choices are possible, like the Davidson potential studied by 
Elliott \cite{Elli} and later by Rowe \cite{Rowe}, that also generates 
an analytic vibration-rotation spectrum. The work of Rowe is very much 
akin to the one discussed here, displaying the same SO(2,1)$\times$SO(5)
algebraic structure.

The purpose of this investigation is to show that,
within the condition of $\gamma-$instability, there 
are other classes of analytically solvable potentials.

The Bohr hamiltonian \cite{Bohr} will be our starting point:
\bea
H\=&-&{\hbar^2\over 2B}\Biggl[{1\over \beta^4}{\partial \over \partial \beta}
\beta^4{\partial \over \partial \beta}+{1\over \beta^2\sin{3\gamma}}
{\partial \over \partial \gamma}\sin{3\gamma}{\partial \over \partial \gamma}
\nonumber \\
~~~&-&{1\over 4\beta^2}\sum_\kappa{Q_\kappa^2 \over 
\sin{\bigl(\gamma-{2\pi\kappa\over 3}\bigr)}^2}\Biggr]+V(\beta,\gamma).
\eea
Following the standard procedure, when the potential
depends only upon $\beta$, i.e. $V(\beta,\gamma)=U(\beta)$, 
one can separate variables as in Wilets and Jean \cite{Wil}, 
obtaining a system of two differential equations, one 
containing the $\gamma$ variable and the three Euler angles, the other 
containing only the $\beta$ variable. The spectrum is determined by the
solution of the latter, namely
\be 
\Biggl\{ 
-{\hbar^2\over 2B}\Biggl[{1\over \beta^4}{\partial \over \partial \beta}
\beta^4{\partial \over \partial \beta} -{\hat \Lambda^2\over \beta^2} 
\Biggr] +U(\beta)\Biggr\} f(\beta) =Ef(\beta)
\label{h2}
\ee
where $\hat \Lambda^2$ is the Casimir operator of SO(5) ($\tau$ being the 
associated quantum number) \cite{Elli} and 
we rewrite here the differential equation in the $\beta$ variable in
the second order standard form as:
\be
\chi''(\beta) +\Bigl\{ \epsilon-u(\beta)-{(\tau+3/2)^2\over \beta^2}+
{1\over 4\beta^2} \Bigl\} \chi(\beta) \=0
\label{diff2}
\ee
where $\chi(\beta)=\beta^{2}f(\beta)$, while $\epsilon={2B\over \hbar^2}E$ 
and $u={2B\over \hbar^2}U$ are the reduced energies and potentials.
In this note we will point out that this equation displays analytic
solutions with the choice of 
the Coulomb and the Kratzer potentials in the quadrupole deformation 
parameter $\beta$.
The former has a very simple form diverging in zero, 
while the
latter has been widely used in the early stages of quantum theory to 
describe interactions within ions in configuration space and has 
a minimum for a finite value of $\beta$.
In both cases we have 
in fact the possibility to regain, with some
 simple mathematical steps, the well-known Whittaker's standard 
form for eq. (\ref{diff2}), and hence to obtain analytic solutions.

\section{Coulomb-like potential}
Inserting the potential
\be
u_C(\beta)\= -{A \over \beta}, \qquad A>0
\ee
in equation (\ref{diff2}) and with the substitutions
$\varepsilon =-\epsilon$,  $x=2 \sqrt{\varepsilon}\beta$, 
$k=A/(2\sqrt{\varepsilon})$ and 
$\mu=\tau+3/2$ the 
previous equation takes the Whittaker's standard form \cite{Bat}:
\be
\chi''(x) +\Bigl\{-{1\over 4}+{k\over x}+{(1/4-\mu^2)\over x^2}
\Bigl\} \chi(x) \=0
\ee
The solution for negative energies, that is regular in the origin,
may be found (as in \cite{Bat,Flu}) to be the Whittaker's function
$M_{k,\mu}(x)$ :
\be
 \chi_{k,\mu}(x)\= {\cal N}_{\tau,\xi}
x^{(2\mu+1)} e^{-x/2} 
~_1F_1 \bigl(\mu+1/2-k,2\mu+1; x\bigr)
\ee
\begin{figure}[t]
\bc
\begin{picture}(280,260)(0,0)
\psset{unit=1.1pt}
\rput(70,0){\small $\xi=0$}
\rput(40,0){\small $\tau$}
\rput(40,10){\small $0$}
\rput(40,130){\small $1$}
\rput(40,172){\small $2$}
\rput(40,191.4){\small $3$}
\rput(40,226){\small $\infty$}
\rput(10,0){\flushleft\small $\varepsilon'$}
\rput(10,10){\flushleft\small 0 }
\rput(10,130){\flushleft\small 1 }
\rput(10,172){\flushleft\small 1.35}
\rput(10,191.4){\flushleft\small 1.512}
\rput(10,226){\flushleft\small  1.8}
\rput(220,120){\small $\xi=1$}
\psline{-}(60,10)(80,10) \rput(60,13){\tiny $0^+$}
\psline{-}(60,130)(80,130) \rput(60,133){\tiny $2^+$}
\psline{-}(60,172)(80,172) \rput(60,175){\tiny $4^+$}
\psline{-}(60,191.4)(80,191.4) \rput(60,194.4){\tiny $6^+$}
\psline{-}(90,172)(110,172) \rput(90,175){\tiny $2^+$}
\psline{-}(90,191.4)(110,191.4) \rput(90,194.4){\tiny $4^+$}
\psline{-}(120,191.4)(140,191.4) \rput(120,194.4){\tiny $3^+$}
\psline{-}(150,191.4)(170,191.4) \rput(150,194.4){\tiny $0^+$}
\psline{-}(210,130)(230,130) \rput(210,133){\tiny $0^+$}
\psline{-}(210,172)(230,172)\rput(210,175){\tiny $2^+$}
\psline{-}(210,191.4)(230,191.4)\rput(210,194.4){\tiny $4^+$}
\psline{-}(240,191.4)(260,191.4)\rput(240,194.4){\tiny $2^+$}
\psline{->}(70,130)(70,10)
\psline{->}(70,191.4)(70,172)
\psline{->}(70,172)(70,130)
\psline{->}(100,172)(72,130)
\psline{->}(100,191.4)(72,172)
\psline{->}(130,191.4)(102,172)
\psline{->}(160,191.4)(104,172)
\psline{->}(100,191.4)(100,172)
\psline{->}(70,191.4)(70,172)
\psline{->}(220,172)(220,130)
\psline{->}(220,191.4)(220,172)
\psline{->}(250,191.4)(221,172)
\psline{->}(220,172)(72,10)
\rput(60,74){\tiny 100}
\rput(60,151){\tiny 878}
\rput(60,186){\tiny 3589}
\rput(95,151){\tiny 878}
\rput(227,151){\tiny 436}
\rput(210,186){\tiny 2478}
\rput(253,186){\tiny 2478}
\rput(140,74){\tiny 2}
\pscircle*[linecolor=white](190,130){15}
\rput(190,120){\small $\tau$}
\rput(190,130){\small $0$}
\rput(190,172){\small $1$}
\rput(190,191.4){\small $2$}
\psset{linestyle=dashed}
\psline{-}(60,226)(260,226)
\end{picture}
\caption{Spectrum of the Coulomb-like potential. 
The energy scale ($\varepsilon'$) is chosen by fixing the 
 energy of the first two states respectively to 0
and 1. The transition rate for $2_{0,1}^+ \rightarrow
0_{0,0}^+$ has been fixed to 100.
Some selected quadrupole transitions are shown in the figure for simplicity. }
\label{Hyd}  
\ec
\end{figure}
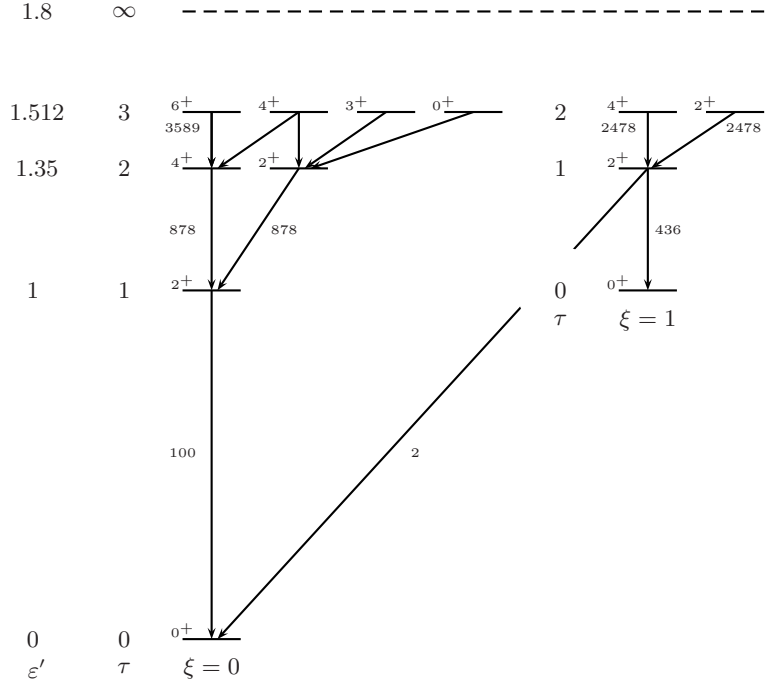
The normalization ${\cal N}_{\tau,\xi}$ of these states can be 
obtained by quadrature 
\footnote{ The general integral of two hypergeometric 
from $0$ to $\infty$ is not known, but one can find the normalization 
constants in an analytical way since the hypergeometrics reduce to 
Laguerre polynomials.}  from:
\be
\int_0^\infty \chi(\beta)^2 d\beta \=\int_0^\infty\beta^4 f(\beta)^2 d\beta  
\=1
\ee
where the volume element in the $\beta$ variable is $\beta^4 d\beta$ 
\cite{Bohr}.
The hypergeometric function for $x\rightarrow \infty$ is in general
proportional to $e^x$ and hence diverges. However, it is not divergent
when it becomes an associated Laguerre
polynomial, that is to say when the first parameter 
$\mu+1/2-k = \tau+2-A/(2\sqrt{\varepsilon})$ is a negative integer, $-\xi$. 
This leads to a condition that fixes the spectrum as
\be
\varepsilon_{\tau,\xi}={A^2/4\over (\tau+\xi+2)^2}
\ee
Now $\tau+\xi$ works as a single quantum number $n$ for the energies, but 
the shape of the wavefunctions depend on $\tau$ and $\xi$ separately.
This spectrum is depicted in figure \ref{Hyd} where
the energy of the first two states have been fixed respectively to 0
and 1 and this is sufficient to settle the energy scale $\varepsilon'$. 
The $(4^+,2^+)$ doublet with $n=2$ has an energy of $1.35$ that is
smaller than the corresponding values for other well-known cases.
Furthermore this spectrum displays another interesting feature: there is 
a threshold at $1.8$ that corresponds to an infinite quantum number.
Again this value is smaller that the energy of the two-phonon state of 
the vibrator.

For the sake of completeness we give the wavefunctions with the 
reduction of the hypergeometric to associated Laguerre polynomials:
\be
\chi(x)_{\tau,\xi}={\cal N}_{\tau,\xi}~ x^{2\mu+1}
e^{-x/2} {\xi!\over (2\mu+1)_\xi} 
L_\xi^{(2\mu)}(x)
\ee
where the denominator is a Pochhammer symbol.

\section{Kratzer-like potential}
\begin{figure}[t]
\bc
\epsfig{file=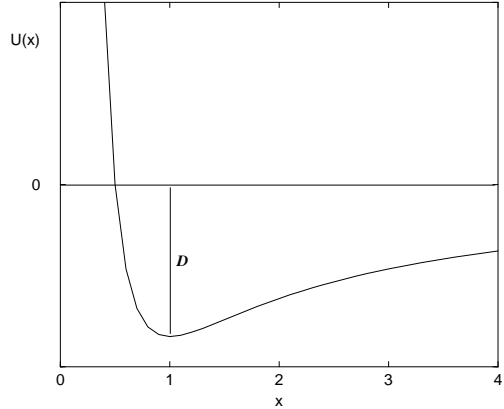,width=0.5\textwidth}
\caption{Kratzer-like potential. For purpose of illustration the horizontal
axis has been set in the adimensional variable $\beta_0/\beta$, 
while the vertical scale is in units of the depth of the pocket.
The minimum corresponds thus to $\beta_0$.}
\label{Kratzer}  
\ec
\end{figure}

We move now to the study of the Kratzer potential (see figure \ref{Kratzer}) 
\be
u_K(\beta)\= -2{\cal D} \Bigl( {\beta_0 \over \beta}-{1\over 2} {\beta_0^2
 \over \beta^2} \Bigr)
\ee
where $\cal D$ represents the depth of the minimum, located in $\beta_0$.
We may also write it as $u_K(\beta)\= u_C(\beta)+B/\beta^2$ for later purpose.
Inserting the cited potential in (\ref{diff2}), with the substitutions 
$\varepsilon =-\epsilon$,  $x=2 \sqrt{\varepsilon}\beta$, 
$k={\cal D}\beta_0/(\sqrt{\varepsilon})$ and 
$\mu^2=(\tau+3/2)^2+{\cal D}\beta_0^2$, the
 equation takes again the Whittaker's standard form.
The regular solutions are again Whittaker's functions with the new 
substitutions and the same arguments apply for the properties of convergence.
Now $\mu+1/2-k = \sqrt{(\tau+3/2)^2+{\cal D}\beta_0^2}+1/2-
{\cal D}\beta_0/\sqrt{\varepsilon}$ must be a negative integer, $-\xi$. 
The spectrum worked out from the last requirement reads: 
\bea
\varepsilon_{\tau,\xi} \= {{\cal D}^2 \beta_0^2 \over (\lambda+\xi)^2}&\=&
{{\cal D}^2 \beta_0^2 \over (\sqrt{(\tau+3/2)^2+ {\cal D}\beta_0^2}+1/2+\xi)^2}
\nonumber \\
~~~~~&\=&{A^2/4 \over (\sqrt{(\tau+3/2)^2+B}+1/2+\xi)^2}
\label{spectrum}
\eea
with $\xi=0,1,2,..$~. The proper set of quantum numbers that characterizes 
the eigenvalues is ${\tau,\xi,L,M}$. Notice that each ${\tau,\xi}$ state 
may be degenerate with respect to the angular momentum, according to the
Wilets and Jean rules \cite{Wil}.  
The $\tau$ quantum number is contained in $\lambda \=
\sqrt{(\tau+3/2)^2+{\cal D}\beta_0^2}+1/2$ and the $\xi$ quantum number
has the same meaning as in the paper of Iachello (shifted by one unity) being
connected with the zeros of the Whittaker's function that are determined by
the zeros of the hypergeometric functions and give rise to the different bands.
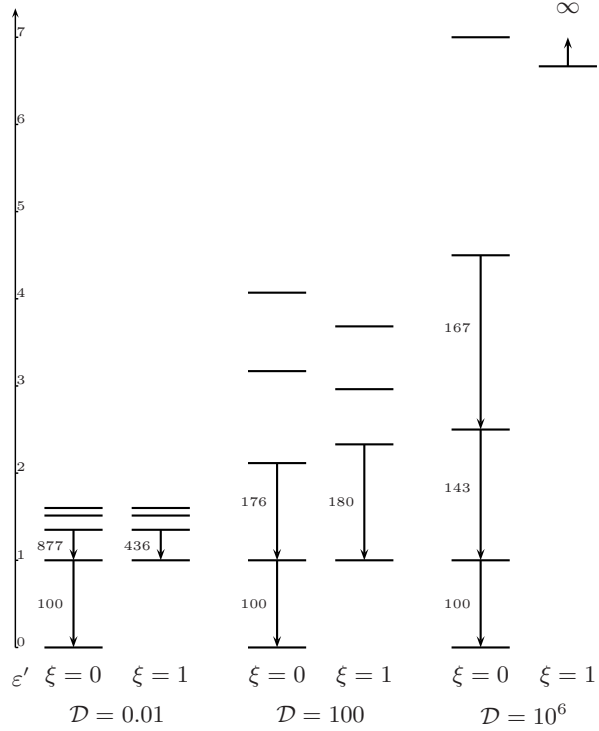
\begin{figure}[t]
\bc
\begin{picture}(220,270)(0,0)
\psset{unit=1.1pt}
\rput(2,10){\small $\varepsilon'$}
\psline[linewidth=0.5]{->}(0,20)(0,240)
\psline[linewidth=0.5]{-}(0,20)(1,20)
\psline[linewidth=0.5]{-}(0,50)(1,50)
\psline[linewidth=0.5]{-}(0,80)(1,80)
\psline[linewidth=0.5]{-}(0,110)(1,110)
\psline[linewidth=0.5]{-}(0,140)(1,140)
\psline[linewidth=0.5]{-}(0,170)(1,170)
\psline[linewidth=0.5]{-}(0,200)(1,200)
\psline[linewidth=0.5]{-}(0,230)(1,230)
\rput(2,22){\tiny 0}
\rput(2,52){\tiny 1}
\rput(2,82){\tiny 2}
\rput(2,112){\tiny 3}
\rput(2,142){\tiny 4}
\rput(2,172){\tiny 5}
\rput(2,202){\tiny 6}
\rput(2,232){\tiny 7}

\rput(20,10){\small $\xi=0$}
\psline{-}(10,20)(30,20)
\psline{-}(10,50)(30,50)       \psline{->}(20,50)(20,20)
\psline{-}(10,60.5)(30,60.5)   \psline{->}(20,60.5)(20,50)
\psline{-}(10,65.4)(30,65.4)
\psline{-}(10,68)(30,68)
\rput(50,10){\small $\xi=1$}
\psline{-}(40,50)(60,50)       \psline{->}(50,60.5)(50,50)
\psline{-}(40,60.5)(60,60.5)
\psline{-}(40,65.4)(60,65.4)
\psline{-}(40,68)(60,68)

\rput(12,35){\tiny 100}
\rput(12,55){\tiny 877}
\rput(42,55){\tiny 436}

\rput(90,10){\small $\xi=0$}
\psline{-}(80,20)(100,20)
\psline{-}(80,50)(100,50)        \psline{->}(90,50)(90,20)
\psline{-}(80,83.45)(100,83.45) \psline{->}(90,83.45)(90,50)
\psline{-}(80,115.1)(100,115.1)
\psline{-}(80,142.1)(100,142.1)
\rput(120,10){\small $\xi=1$}
\psline{-}(110,50)(130,50)
\psline{-}(110,89.9)(130,89.9)   \psline{->}(120,89.9)(120,50)
\psline{-}(110,108.89)(130,108.89)  
\psline{-}(110,130.5)(130,130.5)

\rput(82,35){\tiny 100}
\rput(82,70){\tiny 176}
\rput(112,70){\tiny 180}

\rput(160,10){\small $\xi=0$}
\psline{-}(150,20)(170,20)  \psline{->}(160,50)(160,20)
\psline{-}(150,50)(170,50)  \psline{->}(160,95)(160,50)
\psline{-}(150,95)(170,95)    \psline{->}(160,154.97)(160,95)
\psline{-}(150,154.97)(170,154.97)
\psline{-}(150,230)(170,230)
\rput(190,10){\small $\xi=1$}
\psline{-}(180,220)(200,220)
\psline{->}(190,220)(190,230)
\rput(190,240){\small $\infty$}

\rput(152,35){\tiny 100}
\rput(152,75){\tiny 143}
\rput(152,130){\tiny 167}

\rput(35,-3){\small ${\cal D}=0.01$}
\rput(105,-3){\small ${\cal D}=100$}
\rput(175,-3){\small ${\cal D}=10^6$}
\end{picture}
\caption{Evolution of spectra with fixed $\beta_0=0.5$ and increasing 
${\cal D}$. The first two bands ($\xi=0,1$) are displayed with their lowest
states ($\tau=0,1,2,..$). The various substates are not displayed for the 
sake of simplicity. The transition showed are $4_{2,0}^+\rightarrow 2_{1,0}^+$
and $2_{1,1}^+\rightarrow 0_{0,1}^+$, normalized with $B(E2;2_{1,0}^+
\rightarrow 0_{0,0}^+)=100$.}
\label{evo1}  
\ec
\end{figure}
If one displays the spectrum given in formula (\ref{spectrum}) imposing that 
the ground state $(\tau=0,\xi=0)$ is at zero energy and that the 
$(\tau=1,\xi=0)$ state has $\varepsilon=1$, it is evident that 
one has to play with the
position of the minimum $\beta_0$ and the depth ${\cal D}$ of the potential.
In fig. \ref{evo1} we study the dependence upon the depth of the potential 
well. 
It is seen from the lowest few states of the first two bands that there 
are clearly two limiting cases: when the depth goes to zero the spectrum 
becomes equivalent to the already discussed $1/\beta$ case, that we wish
to call the Coulomb-like limit, while when 
${\cal D}$ tends to infinity all the $\xi>0$ bands escape to an infinite energy
and the $\xi=0$ band has a spectrum that follows a simple $\beta$-rigid, 
$\gamma$-soft rule: $\varepsilon_\tau=\tau (\tau+3)/4$. 
\begin{figure}[t]
\bc
\begin{picture}(220,270)(0,0)
\psset{unit=1.1pt}
\rput(2,10){\small $\varepsilon'$}
\psline[linewidth=0.5]{->}(0,20)(0,240)
\psline[linewidth=0.5]{-}(0,20)(1,20)
\psline[linewidth=0.5]{-}(0,50)(1,50)
\psline[linewidth=0.5]{-}(0,80)(1,80)
\psline[linewidth=0.5]{-}(0,110)(1,110)
\psline[linewidth=0.5]{-}(0,140)(1,140)
\psline[linewidth=0.5]{-}(0,170)(1,170)
\psline[linewidth=0.5]{-}(0,200)(1,200)
\psline[linewidth=0.5]{-}(0,230)(1,230)
\rput(2,22){\tiny 0}
\rput(2,52){\tiny 1}
\rput(2,82){\tiny 2}
\rput(2,112){\tiny 3}
\rput(2,142){\tiny 4}
\rput(2,172){\tiny 5}
\rput(2,202){\tiny 6}
\rput(2,232){\tiny 7}

\rput(20,10){\small $\xi=0$}
\psline{-}(10,20)(30,20) 
\psline{-}(10,50)(30,50)   \psline{->}(20,50)(20,20)
\psline{-}(10,60.5)(30,60.5) \psline{->}(20,60.5)(20,50)
\psline{-}(10,65.4)(30,65.4)
\psline{-}(10,68)(30,68)
\rput(50,10){\small $\xi=1$}
\psline{-}(40,50)(60,50)
\psline{-}(40,60.5)(60,60.5)
\psline{-}(40,65.4)(60,65.4)  \psline{->}(50,60.5)(50,50)
\psline{-}(40,68)(60,68)

\rput(12,35){\tiny 100}
\rput(12,55){\tiny 877}
\rput(42,55){\tiny 436}

\rput(90,10){\small $\xi=0$}
\psline{-}(80,20)(100,20)
\psline{-}(80,50)(100,50)     \psline{->}(90,50)(90,20)
\psline{-}(80,76.01)(100,76.01) \psline{->}(90,76.01)(90,50)
\psline{-}(80,95.51)(100,95.51)
\psline{-}(80,109.5)(100,109.5)
\rput(120,10){\small $\xi=1$}
\psline{-}(110,68.9)(130,68.9)
\psline{-}(110,84.8)(130,84.8)   \psline{->}(120,84.8)(120,68.9)
\psline{-}(110,99.65)(130,99.65)
\psline{-}(110,111.5)(130,111.5)

\rput(82,35){\tiny 100}
\rput(82,63){\tiny 222}
\rput(112,76){\tiny 212}

\rput(160,10){\small $\xi=0$}
\psline{-}(150,20)(170,20)    \psline{->}(160,50)(160,20)
\psline{-}(150,50)(170,50)   \psline{->}(160,95)(160,50)
\psline{-}(150,95)(170,95)     \psline{->}(160,154.97)(160,95)
\psline{-}(150,154.97)(170,154.97)
\psline{-}(150,230)(170,230)
\rput(190,10){\small $\xi=1$}
\psline{-}(180,220)(200,220)
\psline{->}(190,220)(190,230)
\rput(190,240){\small $\infty$}

\rput(152,35){\tiny 100}
\rput(152,75){\tiny 143}
\rput(152,130){\tiny 167}

\rput(35,-3){\small $\beta_0=0.01$}
\rput(105,-3){\small $\beta_0=1$}
\rput(175,-3){\small $\beta_0=100$}
\end{picture}
\caption{Evolution of spectra with fixed ${\cal D}=10$ and increasing 
$\beta_0$. The first two bands ($\xi=0,1$) are displayed with their lowest
states ($\tau=0,1,2,..$). The various substates are not displayed for the 
sake of simplicity. The selected transition rates are the same than in the 
preceding figure.}
\label{evo2}  
\ec
\end{figure}
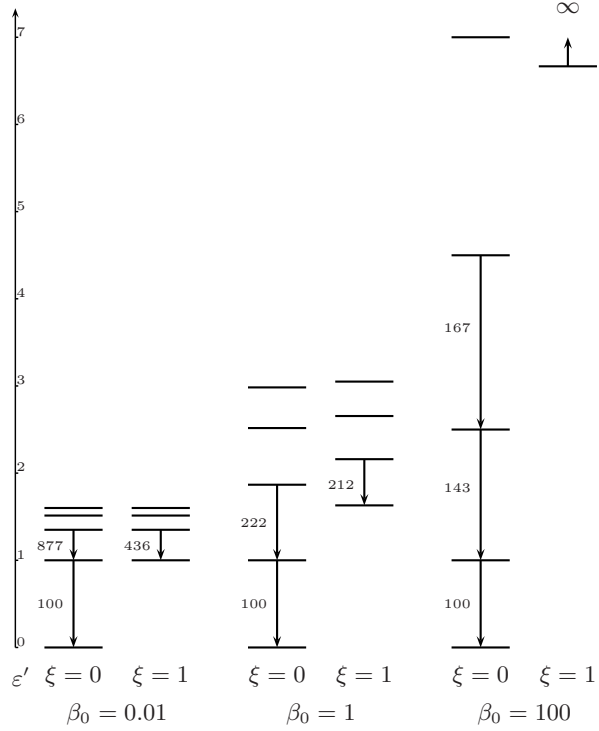
In fig. \ref{evo2} we study instead the evolution of the spectra with $\beta_0$
for a fixed value of ${\cal D}$. Again the same two limits are seen when 
the deformation parameter is very small or very large. 
There is a qualitative equivalence between the case with small $\beta_0$,
 the case with small ${\cal D}$ and the Coulomb-like case. Moreover when
$\beta_0$ or ${\cal D}$ tend to infinity the situation is equivalent.\\
Apart from the limiting cases the most interesting situations are realized 
for spectra that have a reasonable $\beta_0$ (that is usually smaller than 0.5)
and a ${\cal D}$ freely adjustable that can widen the relative distances in
the $\xi=0$ band as well as move the lowest state of the second $\xi=1$ band.

\section{Transitions}
Electromagnetic transition rates (see \cite{Iac1,Cap,Aria}) are 
defined as reduced matrix elements of the transition operator $T(E\lambda)$
\be
B(E\lambda;\xi_i,\tau_i,L_i\rightarrow \xi_f,\tau_f,L_f)={
\mid \langle \xi_i,\tau_i,L_i \| T(E\lambda) \|  
\xi_f,\tau_f,L_f \rangle\mid^2 \over (2L_i+1)}
\ee
and the quadrupole transition operator, to the first order, reads:
\be
T(E2,\mu)\propto \beta \left[ D_{\mu,0}^{(2)}\cos{\gamma}+
(D_{\mu,2}^{(2)}+D_{\mu,-2}^{(2)}){\sin{\gamma}\over \sqrt{2}}   \right]
\ee
The matrix elements have been calculated for a few selected transitions
and are displayed directly on the figs. \ref{Hyd},\ref{evo1} and \ref{evo2}.
The $\beta$ part of the integration factorizes out and the 
$\{\gamma,\theta_i \}$ part has been calculated in the standard way.
Taking into account all the transitions between
a given $\tau$ state and one of his neighbouring (for instance $\tau-1$)
states, it suffices to calculate explicitly one transition, 
the others being obtained by fixed geometrical factors coming 
from the $\gamma$-angular part \cite{Bes}.
It is interesting to note that the transition from the first $2+$ state of
the second band to the ground state of the system is very small if compared
to the reference transition, evidencing almost no overlap  between
 the two wavefunctions in the $\beta$ variable.

For simplicity in the last two figures we have given only the 
following transition rates: $6_{3,0}^+\rightarrow 4_{2,0}^+$, 
$4_{2,0}^+\rightarrow 2_{1,0}^+$ and  $2_{1,1}^+\rightarrow 0_{0,1}^+$,
relatively to the transition $2_{1,0}^+\rightarrow 0_{0,0}^+$
that has been given the reference value of 100.
It is interesting to note that in both cases (fixing one parameter and 
changing the other) the two evidenced limits have also common values
for B(E2)'s and the two Coulomb-like limits for the Kratzer potential 
show the same values of the exact Coulomb-like case in fig. \ref{Hyd}.

\section{Algebraic structure}
We discuss now the group theoretical interpretation of
the hamiltonians that we have solved directly in the previous sections.
We will consider the quantization of the collective model, restricted to
quadrupole deformations only ($\lambda=2$),
by defining pairs of canonically conjugate operators on the Hilbert space
$\hat \alpha_\mu \=\alpha_\mu$ and $\hat \pi^\mu\=-i{\partial\over \partial 
\alpha_\mu}$ (having dropped any $\hbar$ and using a covariant/contravariant 
notation). These operators obey the Heisenberg-Weyl commutation relations 
$[ \hat\alpha_\mu,\hat\pi^\nu ] \=i\delta_{\mu,\nu}$ \cite{Rowe2}.
If $\hat {\vec \alpha}$ is defined as the vector whose five components are the 
operators defined above, the scalar product may be written as 
\be
\hat {\vec \alpha} \cdot \hat{\vec \alpha} \= 
\sum_\mu \alpha^\mu\alpha_\mu \=\sum_\mu \mid \alpha^\mu \mid^2
\ee
where the last equivalence is a consequence of the property: $\alpha^\mu\=
\alpha_\mu^*\=(-1)^\mu\alpha_{-\mu}$.
Notice that $\beta^2 \= \hat {\vec \alpha} \cdot \hat {\vec \alpha}$.
We need also to consider the parameters $a_\mu$ and $p_\mu$ 
which play the role of the $\alpha_\mu$ and 
$\pi_\mu$ in the intrinsic frame of reference, and have the useful property
that $a_\mu \=a_{-\mu}$. Since the transformations between
these two sets of variables are unitary, the commutation relations are 
preserved as well as the scalar products  
(for example $\sum_\mu \pi^\mu \pi_\mu\=\sum_\mu p^\mu p_\mu$).

Taking the reduced quantities, equation (\ref{h2}) may be recast in the form
\be
\Bigl[ \underbrace{\pi^2+u(\beta)}_{\cal H} -\epsilon\Bigr] f(\beta)=0
\ee
where (see \cite{Elli})
\be
\pi^2\=\hat{\vec \pi}\cdot\hat{\vec \pi}\=\hat{\vec p}\cdot\hat{\vec p} 
\= -{1\over \beta^4}{\partial \over \partial \beta}
\beta^4{\partial \over \partial \beta} +{\hat \Lambda^2\over \beta^2}.
\ee

For both the Coulomb-like and Kratzer-like 
potentials the hamiltonian ${\cal H}$ is SO(5) invariant.
In addition it displays a spectrum generating algebra since one may 
construct three operators \cite{Rowe} that are infinitesimal generators 
for the corresponding group: 
\be
\hat Z_1\=4\beta\Bigl(p^2+{B\over\beta^2}\Bigr) \qquad ~
\hat Z_2\=\beta \qquad ~
\hat Z_3\=2\bigl(\hat{\vec a}\cdot \hat{\vec p} -i\bigr)
\ee
which have the commutation relations of the four non-compact isomorphic Lie 
algebras su(1,1) $\sim$ so(2,1) $\sim$ sl(2,R)$\sim$ sp(2,R) (using 
Wybourne's notation for symplectic group dimensions \cite{Wyb}):
\be
[\hat Z_1,\hat Z_2]=-4i \hat Z_3 \qquad ~
[\hat Z_3,\hat Z_2]=-2i \hat Z_2 \qquad ~
[\hat Z_3,\hat Z_1]=2i \hat Z_1.
\ee 
With the potential $u(\beta)=u_K(\beta)$ (that contains also the 
Coulomb-like case, when $B=0$)
the operator $\beta{\cal H}$ is in fact expressible as a linear combination 
of the elements of the algebra of $su(1,1)$, namely in the form 
\be
\beta {\cal H} \=\hat Z_1/4-A.
\label{bH}
\ee
Now one can define new operators ${\hat X_i}$ with $i=1,2,3$ by means of a 
linear transformation:
\be
\hat X_1\={1\over 4} \bigl( \hat Z_1-\hat Z_2 \bigr) \qquad ~
\hat X_2\={1\over 2} \hat Z_3 \qquad ~
\hat X_3\={1\over 4} \bigl( \hat Z_1+\hat Z_2 \bigr)
\ee
which satisfy the following commutation relations:
\be
[\hat X_1,\hat X_2]=-i \hat X_3 \qquad ~
[\hat X_2,\hat X_3]=i \hat X_1 \qquad ~
[\hat X_3,\hat X_1]=i \hat X_2.
\ee 
The eigenvalue equation for the Bohr hamiltonian is now, having 
multiplied by $\beta$ on the left, 
\be
\beta \bigl( {\cal H}-\epsilon \bigr)\Psi\=0.
\ee
Using equation (\ref{bH}) and the definitions of the $\hat X_i$ and 
$\hat Z_i$ operators, we rewrite the last equation as
\be
\Bigl[ (1+4\epsilon)\hat X_1 -2A + (1-4\epsilon)\hat X_3 \Bigr] \Psi \=0
\ee
and following the procedure in \cite{Ald} we can perform a (1,3) 
hyperbolic rotation of an angle $\theta$
to diagonalize the eigenvalue equation and choosing $tgh(\theta)= - (1+4
\epsilon)/(1-4\epsilon)$ (valid for $\epsilon <0$) we obtain
\be
\hat X_3 \tilde\Psi \= {A\over \sqrt{-4\epsilon}} \tilde \Psi,
\ee
where $\tilde \Psi$ is the rotated wavefunction.
The Casimir operator of the so(2,1) algebra is evaluated to be:
\be
\hat \mathbb{C}_2\= \hat \Lambda^2 + \hat X_- - \hat X_+ + B + 2 
\ee
with eigenvalue $\tau(\tau+3)+B+2$.
The two last equations must be  compared with the two following eigenvalue
equations (for unitary representations $D^+$ \cite{Baru}):
\bea
\hat X_3 \mid \phi,\xi\rangle &\=& (\xi-\phi)\mid \phi,\xi\rangle\nonumber \\
\hat \mathbb{C}_2 \mid \phi,\xi\rangle &\=& \phi(\phi+1)\mid \phi,\xi\rangle.
\eea
This comparison yields a spectrum of the form:
\be
\epsilon_{\tau,\xi} \=-{A^2/4 \over (\sqrt{(\tau+3/2)^2+B}+1/2+\xi)^2}
\ee
that coincides with the one found from the direct solution of the differential 
equation with a Kratzer-like potential. It also contains as a special 
case ($B=0$) the Coulomb-like case. 

The algebra associated with the SO(5) group is the so-called degeneracy 
algebra (according to the definitions in \cite{Cord}) and the group 
SO(2,1) is associated with the spectrum generating algebra.
The relevant chain of subalgebras that gives the labels of the set of 
orthonormal states $\{\mid \xi \tau \alpha L M \rangle \}$ is explicitly
given as
\cite{Rowe,Rowe2}:
\be
\begin{array}{ccccccccc}
SU(1,1)&\times &SO(5)&\supset& U(1)&\times & SO(3)&\supset & SO(2)\\
\lambda&~& \tau &\alpha &\xi&~&L&~&M \\
\end{array}
\ee
where $\lambda$ is an SU(1,1) lowest weight and $\alpha$ indexes the SO(3)
multiplicity. These basis diagonalize the hamiltonian (\ref{bH}) given above. 
We can thus state that the problem studied so far displays a 
SO(2,1)$\times$SO(5) dynamical algebra.

\section{Final remarks}
In this paper we have solved the Bohr hamiltonian for two specific 
$\gamma-$unstable potentials that yield analytical solutions.
We have given the corresponding spectra and wavefunctions in closed form
and the most important transition probabilities.
Rather interesting looks the case of the Kratzer-like potential 
that may be given in terms of two unrelated parameters $A$ and $B$, 
as $-A/\beta+B/\beta^2$.
Changing the value of $B$ from zero to a finite value, one can describe
situations ranging from spherical to
 quadrupole deformed shape. The critical point is actually the potential with 
$B\rightarrow 0$, that is the Coulomb-like limit discussed above. 
The same dynamical symmetry discussed briefly in the former section,
is thus effective for a class of different potentials, that, depending on 
some parameter, may describe very different situations.
We would like to remark that these potentials parallel the case of 
the family of potentials of the form: 
\be
u_H(\beta)=A\beta^2
\ee
and
\be 
u_D(\beta)=u_H(\beta)+{B\over \beta^2}
\ee
where $H$ stays for harmonic and $D$ for Davidson, that also lead to 
solvable Bohr's
hamiltonians and are furthermore both characterized by a 
SO(2,1)$\times$SO(5) dynamical group \cite{Rowe}. 
Other interesting possibilities, such as linear combinations of powers (to
be solved with Frobenius method), arise in the same spirit of this paper.

The $\beta-$part of the spectrum of the potentials that have been 
discussed here was
combined with the condition of $\gamma-$instability. We will show in a 
forthcoming paper that approximate solutions can also be obtained for the
same functional dependence on $\beta$, but with a $\gamma-$dependence 
favouring axial symmetry. 
This case will be therefore the homologous of the situation associated with
the occurrence of X(5) in the approximate solution of reference
\cite{Iac2}.

\section*{Acknowledgments}
We wish to acknowledge valuable correspondence and discussions with
J.M.Arias, F.Iachello and especially with D.J.Rowe.

\end{document}